\documentclass[runningheads]{llncs}

\usepackage{enumitem}
\usepackage{makecell}
\usepackage{float}
\usepackage{booktabs}
\usepackage{listings,xcolor}
\newfloat{lstfloat}{htbp}{lop}
\floatname{lstfloat}{Listing}

\definecolor{dkgreen}{rgb}{0,.6,0}
\definecolor{dkblue}{rgb}{0,0,.6}
\definecolor{dkyellow}{cmyk}{0,0,.8,.3}

\lstdefinestyle{php}{
	language        = php,
	basicstyle      = \footnotesize\ttfamily,
	keywordstyle    = \color{dkblue},
	stringstyle     = \color{red},
	identifierstyle = \color{dkgreen},
	commentstyle    = \color{gray},
	emph            =[1]{php},
	emphstyle       =[1]\color{black},
	emph            =[2]{if,and,or,else},
	emphstyle       =[2]\color{dkyellow},
	numbers=left,
	firstnumber=1,
	numberfirstline=true
	numberstyle=\tiny,
	numbersep=-1pt,
	tabsize=1,
	extendedchars=true,
	breaklines=true,
	frame=lines,
	showspaces=false,
	showtabs=false,
	showstringspaces=false,
	breakatwhitespace=false,
	showlines=true,
	captionpos=b,
}

\lstdefinestyle{nocoloring}{
	keywordstyle=\color{black},
	commentstyle=\color{black},
	stringstyle=\color{black},
	identifierstyle = \color{black},
	basicstyle      = \footnotesize\ttfamily,
	numbers=left,
	firstnumber=1,
	numberfirstline=true
	numberstyle=\tiny,
	numbersep=-1pt,
	tabsize=1,
	extendedchars=true,
	breaklines=true,
	frame=lines,
	showspaces=false,
	showtabs=false,
	showstringspaces=false,
	breakatwhitespace=false,
	showlines=true,
	captionpos=b,
}

\usepackage[T1]{fontenc}
\usepackage{graphicx}
\PassOptionsToPackage{hyphens}{url}\usepackage{hyperref}
\usepackage{color}

\usepackage{cleveref}
\usepackage{amssymb}

\newcommand{\FUEL}{\textsc{FUEL}}

\begin{document}
\title{Bringing UFUs Back into the Air With \FUEL:\\ A Framework for Evaluating the Effectiveness of Unrestricted File Upload Vulnerability Scanners}
\titlerunning{\FUEL: A framework to evaluate UFU Vulnerability Scanners}
%
\author{Sebastian Neef\inst{1}\orcidID{0000-0003-3055-0823} \and
	Maath Oudeh\inst{1}\orcidID{0009-0007-9131-2200}}
\authorrunning{S. Neef and M. Oudeh}
%
\institute{Security in Telecommunication, Technische Universität Berlin, Germany\\
	\email{\{neef,maath.oudeh\}@tu-berlin.de}\\}
\maketitle              

\begin{abstract}
Unrestricted file upload (UFU) is a class of web security vulnerabilities that can have a severe impact on web applications if uploaded files are not sufficiently validated or securely handled.
A review of related work shows an increased interest in finding new methods to discover such vulnerabilities. 
However, each publication evaluates its new vulnerability scanner against a different set of artificial or real-world applications available at the time of writing.
Thus, we identify the need for a comprehensive testing framework to allow a reproducible comparison between existing and future UFU vulnerability scanners.

Our contributions include the File Upload Exploitation Lab (\textit{\FUEL}), which models 15 distinct UFU vulnerabilities in isolated scenarios to enable a reproducible evaluation of UFU scanners' capabilities.
The results of evaluating four black-box UFU scanners against FUEL show that no scanner manages to identify all UFU vulnerabilities, leaving real-world websites at risk of compromise due to false negatives. 
Our work aims to solve this problem by extending an existing UFU scanner with multiple new detection and exploitation techniques, which we call \textit{Fuxploider-NG}, to increase its accuracy from $\sim$50\% to over 90\%, thereby surpassing the capabilities of existing UFU scanners and showcasing the importance of \FUEL~as a UFU vulnerability evaluation framework.
To foster open science and future work in this area, we open-source \FUEL~and Fuxploider-NG.

\end{abstract}
\section{Introduction}
The World Wide Web (WWW) is constantly evolving, offering its users the possibility to contribute content since the release of the web 2.0 \cite{jacksi2019development}.
According to \cite{internetusers}, recent statistics count over 5.3 billion internet users, with a large quantity being social media users. 
Websites, including social media platforms, allow users to contribute content in various ways, e.g., by uploading files such as documents, images, videos, or other types. 
Sharing files through web applications has become integral for businesses, social networks, personal websites, or other domains that rely on web applications.

Despite its 30th anniversary, the Hypertext Preprocessor (PHP) programming language remains the most popular for web development, with a share of over 75\% based on statistics from \cite{php-popularity}, powering websites such as facebook.com or wikipedia.org \cite{php-websites}.
Modern web servers such as Apache httpd or NGINX provide extensions to execute PHP code, further aiding its popularity \cite{apache-php,nginx-php}.
For example, the PHP-based content management system WordPress claims to have powered over 40\% of the web in 2021 \cite{wordpress_popularity}.

PHP supports processing uploaded files with special variables, functions, and configuration options \cite{php-files,php-postfile}.
However, it is crucial to handle uploaded files carefully and securely, or unrestricted file upload (UFU) vulnerabilities may arise \cite{owasp-ufu}.
Especially in PHP-developed web applications, failure to properly validate uploaded files may result in full system compromise (e.g., remote code execution) or other high-impact attacks (e.g., cross-site scripting).

File upload vulnerabilities are prevalent to this day.
Public vulnerability databases, such as \cite{exploitdbOffSecsExploit}, list over 1,000 public exploits in web applications related to UFU.
Similarly, the Common Vulnerability Enumeration (CVE) database contains over 3,300 entries matching the keyword ``file upload'' \cite{mitreSearchResultsFileUpload}.
Furthermore, WPScan documents several thousand security issues, including many authenticated and unauthenticated UFU vulnerabilities, in WordPress and its third-party plugins or themes \cite{wpscan-vulns}.

The academic community has shown great interest in researching UFU vulnerability detection methods and contributing automated vulnerability scanners, as detailed in \cref{sec:related-work}.
Additionally, commercial (e.g., BurpSuite \cite{portswigger-burp}) and open-source (e.g., Fuxploider \cite{fuxploider} or ZAP \cite{owasp-zap}) scanners have emerged and are being actively developed to help identify UFU vulnerabilities.

However, none of the academic publications have used a standardized method for evaluating their new approaches but have used artificial or real-world PHP-based web applications available at the time of writing.
Our results will show that the lack of a standardized evaluation environment has led to shortcomings in all UFU vulnerability scanners' detection capabilities.
Consequently, users who trust the output of the currently available UFU scanners might assume their web application to be secure against UFU vulnerabilities, although the missing detection capabilities and false-negatives might allow attackers to still compromise the application.

Motivated by these discoveries, we aim to extend the research body in the UFU vulnerability detection domain with the following contributions:
\begin{itemize}
    \item We present \FUEL~as an extensible and modular testing framework to facilitate reproducible and comprehensive evaluations of UFU scanners.
    \item We evaluate four academic, open-source, and commercial UFU scanners against a set of 15 unique UFU scenarios implemented in \FUEL~that reveals many shortcomings in these scanners.
    \item We implement \textit{Fuxploider-NG}, which consolidates and surpasses the capabilities of the evaluated vulnerability scanners into a single tool.
\end{itemize}

\section{Background \& Related Work}
Although the reader is assumed to be familiar with general concepts related to web technologies, e.g., the Hyper Text Transfer Protocol (HTTP), Hypertext Markup Language (HTML), or common web vulnerabilities, e.g., the Open Worldwide Web Application Project (OWASP) Top 10 \cite{owaspTopTen}, this section provides an overview of UFU related knowledge and related work to follow the remainder of this work.

\subsection{PHP}
The Hypertext Preprocessor (PHP) programming language was specifically designed for web development in 1994 \cite{php-history} and has continuously evolved to reach version PHP 8.3 at the time of writing.
We assume one reason for its high adoption is its simplicity, as HTML and PHP code can be mingled to create basic, dynamic web applications, e.g., \texttt{<h1><?php echo \$\_GET['name'];?></h1>}.
To execute PHP code, a PHP file must be passed to the PHP interpreter, which is performed by modern web servers such as Apache, NGINX, or others \cite{apache-php,nginx-php}.
Thus, the barrier to entry for developing feature-rich websites using PHP is low, and vulnerabilities, including UFU, might arise from simple programming mistakes \cite{owaspTopTen}.

\subsection{File Upload Process}
\begin{figure}[t]
\centering
\includegraphics[width=0.69\textwidth]{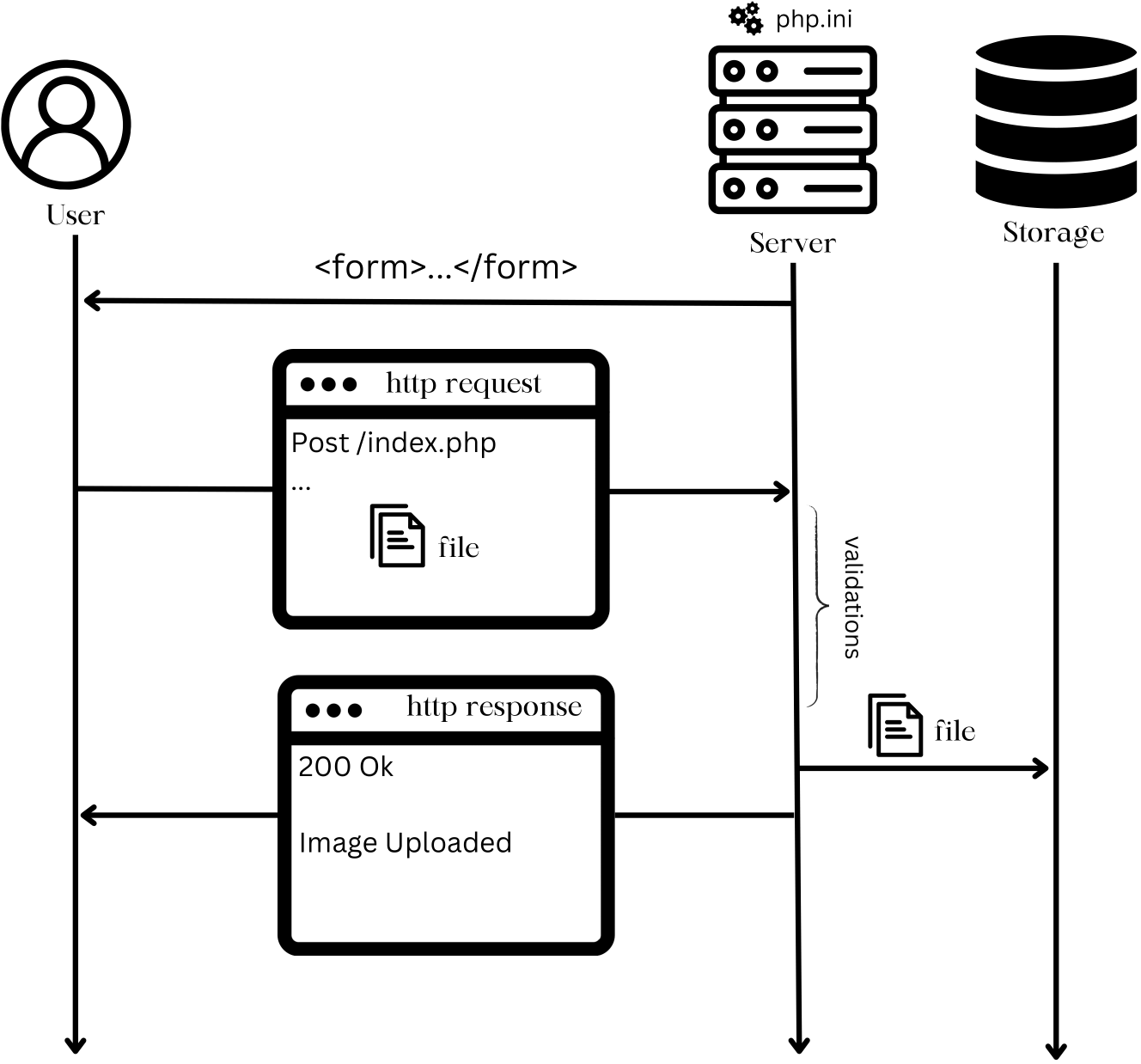}
\caption{A simplified schema of the file upload and retrieval steps.}
\label{fig:file-upload-process}
\end{figure}
Understanding the fundamental steps in a file upload process is crucial for discovering UFU vulnerabilities.
While the client-side handling of file uploads, e.g., letting the user choose a file and initiating the file transfer, is part of the HTML (e.g., \cite{w3_fileapi,w3_fileinput}) and HTTP (e.g., \cite{RFC7578,RFC1867}) specifications implemented in the browser, the server-side handling can differ based web application's implementation and web server's configuration. 
\Cref{fig:file-upload-process} provides a simplified depiction of the basic components and interactions of a web-based file upload process. 

\subsubsection{File Selection}
First, the web server sends the user a website containing an HTML form with an input field of type ``file'', as shown in \cref{lst:file-upload-form}.
The form's attributes inform the browser how to handle the file upload, e.g., the HTTP request method in \texttt{method}, the web server's receiving endpoint in \texttt{action}, and the form data's encoding in \texttt{enctype}. 
With the help of additional HTML properties or JavaScript, the user's file selection can be restricted and validated in the browser.
However, these client-side restrictions can be circumvented, e.g., by intercepting and modifying the resulting HTTP request.
\begin{lstfloat}[t]
    \begin{lstlisting}[caption={A basic HTML file upload form.},label={lst:file-upload-form},style=php]
 <form enctype="multipart/form-data" method="POST" action="/index.php">
     <input type="file" name="fileToUpload"/>
     <input type="submit" value="Upload" name="submit"/>
 </form>
    \end{lstlisting}
\end{lstfloat}

\subsubsection{File Transfer}
\begin{figure}[t]
\centering
\includegraphics[width=0.8\textwidth]{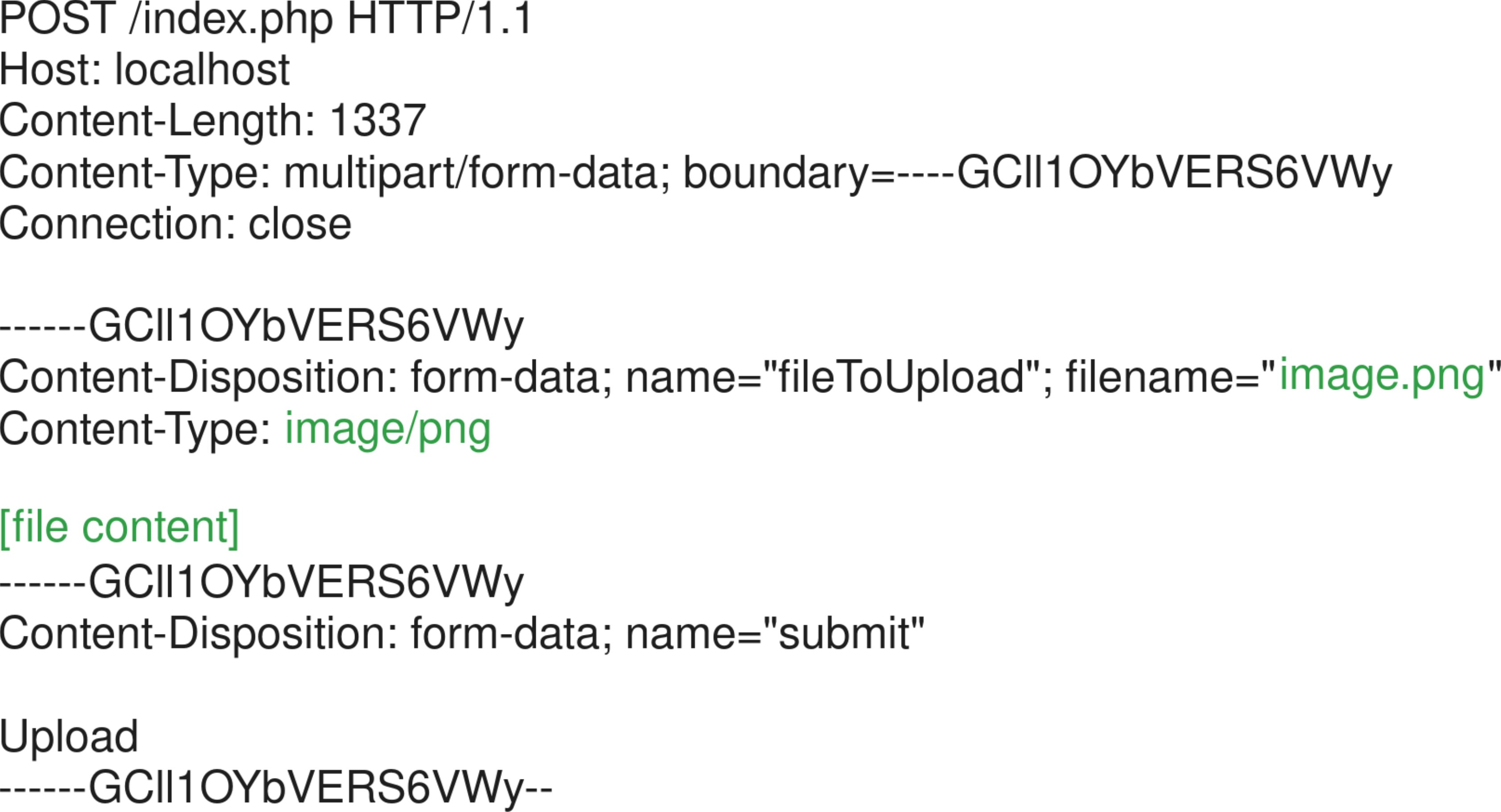}
\caption{An example \texttt{multipart/form-data} encoded file upload HTTP request with the UFU-relevant manipulation points highlighted.}
\label{fig:http-upload-req}
\end{figure}

After the user has chosen a file to upload and submitted the form, the browser initiates the file transfer.
RFC 7578 specifies the content type used by file uploads to be \texttt{multipart/form-data} \cite{RFC7578}.  
\Cref{fig:http-upload-req} shows the HTTP request issued by the browser to the server.
In order to include the file's data in the request, the \texttt{multipart/form-data} encoding transforms all form fields into blocks separated by a randomly generated boundary and a \texttt{Content-Disposition} line.
Additionally, the request contains information about the file, e.g., the name, content, and MIME type (cf. RFC 2045ff. \cite{RFC2045}).

\subsubsection{File Processing}
PHP processes uploaded files by default and offers additional configuration options to restrict incoming file uploads \cite{php-uploads}, e.g., setting a maximum file size.
Generally, uploaded files are assigned a random name and stored temporarily in a specifically configured directory.
The PHP application accesses the files and their metadata through a 2D associative array named \texttt{\$\_FILES} \cite{php-files,php-postfile}.
The first dimension's keys are the file-fields' names, e.g., \texttt{fileToUpload}, and the second dimension provides a file's properties as shown in \cref{lst:php-files-array}.
While the key \texttt{name} represents the file's name on the client's file system, \texttt{tmp\_name} is the random file name generated by PHP.

The attacker controls the \texttt{name} and \texttt{type} properties as well as the file's content.
Therefore, web applications must correctly validate these to avoid UFU vulnerabilities \cite{owasp-ufu}.
As the next section clarifies, the example provided in \cref{lst:php-files-array} is not secure.
After validation, the web application may eventually wish to store the file by moving it from its temporary location to a permanent location, e.g., using \texttt{move\_uploaded\_file}.

\begin{lstfloat}[b]
    \begin{lstlisting}[caption={A basic example of insufficient file property validation in PHP},label={lst:php-files-array},style=php]
 $fname = $_FILES["fileToUpload"]["name"];
 $ftype = $_FILES["fileToUpload"]["type"];
 $fsize = $_FILES["fileToUpload"]["size"];
 $ftmp  = $_FILES["fileToUpload"]["tmp_name"];
 if($fsize < 100*1024 && $ftype = "image/png") {
     move_uploaded_file($ftmp, "/uploads/" . $fname);
 }
    \end{lstlisting}
\end{lstfloat}

\subsection{Unrestricted File Upload}\label{sec:file-upload-issues}
Although the actual risks and consequences of an insecure file upload process are dependent on several factors, e.g., its implementation, the web server's or storage layer's configuration, UFU vulnerabilities root in insufficient validation and typically result in code execution, cross-site scripting, or other high consequence attacks \cite{owasp-ufu}.

\subsubsection{Improper Validation}
The failure to validate attacker-controlled file properties may result in the upload of dangerous files \cite{pooj2016understanding,owasp-ufu}.

Failure to correctly validate the file name may allow the attacker to create files in arbitrary storage locations, thus overwriting existing files or injecting HTML markup into the web application.
The code in \cref{lst:php-files-array} performs no file name validation whatsoever.

Incorrect file type validation might allow the attacker to trick
the application into assuming a permissible file was uploaded, although the content is malicious.
For example, the validation in \cref{lst:php-files-array} assumes a \texttt{image/png} type but does not ensure that the type matches the file content. 
Attackers may use polyglot files \cite{github-polyglots}, which are combinations of two or more file types into a single file, to bypass insufficient file type or file content checks.

Thus, secure web applications must verify that the file content matches the file's type to avoid confusion.
However, examining the content is a complex task that depends on the file type or size, as large files may require more processing resources and time.
Furthermore, attackers might exploit vulnerabilities in the parsers, libraries, or other components used for validation, e.g., ImageTragick \cite{imagetragick}. 

Finally, failing to verify a file's size may result in large files being uploaded repeatedly, thus eventually exhausting the storage space.
Depending on the web server's storage configuration, this may lead to consequent HTTP requests or uploads to fail, rendering the website unresponsive.

\subsubsection{Vulnerabilities}
Although the vulnerabilities resulting from insecure file uploads can be manifold \cite{owasp-ufu}, our evaluation focuses on code execution and cross-site scripting, similar to related work (cf. \cref{sec:related-work}), as these have a high impact.

In the context of PHP web applications, several methods exist to achieve \textit{Code Execution (CE)} against the web application through UFU.
For example, one method is uploading a file containing PHP code and instructing the PHP interpreter to execute it, e.g., uploading \texttt{exploit.php} and subsequently accessing it.
Other methods include changing the file's content to exploit outdated or vulnerable components that process the uploaded file, such as ImageTragick \cite{imagetragick}, or uploading configuration files that change the interpreter's or web server's behavior, e.g., \texttt{.htaccess} files \cite{apache-htaccess}.

In contrast to code execution, \textit{Cross-Site Scripting (XSS)} attacks target the website's users.
Insufficient validation might allow an attacker to upload \texttt{.html},\texttt{.htm}, or other files containing XSS payloads in a way that the web server serves these with a \texttt{text/html} content type. 
Thus, a victim's browser rendering such a file will interpret the HTML markup and execute JavaScript payloads.
Another conceivable attack is uploading files with an XSS payload in the file name, which is then insecurely emitted by the web application in its HTML context.

\vfill
\subsection{Related Work}\label{sec:related-work}

File upload vulnerabilities are not a new research topic. 
First mentions of UFU issues can be traced back to 2009 when Barth et al. discussed content sniffing attacks \cite{barth2009secure} and 2011 when Barua et al. evaluated server-side content sniffing detection \cite{barua2011server} against PhpBB2, Prestashop, and Serendipity.

Dahse et al. used six different PHP projects, including osCommerce, to evaluate their method to detect second-order vulnerabilities, including UFU, in 2014 \cite{dahse2014static}.
In 2016, Riadi et al. analyzed \textit{osCommerce} for UFU vulnerabilities \cite{riadi2016analysis}, while Pooj et al. compiled 16 different file upload vulnerability cases.
In 2017, De Meo et al. \cite{wafex} used formal methods to detect UFU in DVWA.

Two years later, Huang et al. \cite{huang2019uchecker} developed \textit{UChecker} to detect UFU vulnerabilities automatically and evaluated it on over 9,000 WordPress plugins.
In 2021, they again used over 9,000 WordPress plugins to evaluate their new scanner \textit{UFuzzer} \cite{huang2021ufuzzer}.
Lee et al. proposed the vulnerability scanner \textit{FUSE} dedicated to finding UFU vulnerabilities in 2020 and evaluated it on a variety of 33 real-world PHP applications from different sources \cite{lee2020fuse}.

In 2022, Yenduri et al. discussed vulnerabilities in PHP and their prevention, including UFU vulnerabilities, and implemented two custom PHP web applications for their evaluation \cite{yenduri2022php}.
The same year, Wichmann et al. \cite{wichmann2022fileuploadchecker} evaluated \textit{FileUploadChecker} against WordPress, Drupal, XWiki, and OpenCMs.
The most recent publication is by Chen et al. proposing \textit{URadar}'s adaptive dynamic testing approach, which they evaluate on 18 PHP web applications from over three different sources \cite{chenUradar2023}.

Over time, several intentionally vulnerable web applications were developed, which also encompass UFU vulnerabilities:
\textit{WackoPicko} simulated a real-world application and was published by Doup{\'e} et al. in 2010 for a study about black-box vulnerability scanners \cite{doupe2010johnny}.
In 2013, the \textit{Damn Vulnerable Web Application} (DVWA) started development to become a PHP-based vulnerability training platform featuring multiple difficulty levels \cite{github-dvwa}.
In 2014, the \textit{JuiceShop} launched as a deliberately vulnerable online shop covering a plethora of vulnerability classes \cite{owasp-juiceshop}.
In contrast to DVWA or WackoPicko, the JuiceShop is developed in NodeJS, which leads to disparities in how uploaded files become dangerous.
All three vulnerable web applications focus on a broad spectrum of vulnerability classes, thus having a limited amount of UFU scenarios, which we assume to be one reason for their seldom reference in related work.

Our literature review revealed that no standardized framework exists to evaluate UFU vulnerability scanners.
Instead, each study chose its own set of artificial or real-world PHP applications, available at the time of writing, to evaluate their novel methodology or scanner.
While this enables the discovery of previously unknown vulnerabilities in existing software, it renders a just comparison of the capabilities and improvements of vulnerability scanners infeasible.
As our evaluation in \cref{sec:evaluation} will show, this leads to several areas for improvement in the capabilities of the examined UFU vulnerability scanners.
\section{FUEL}\label{sec:fuel}
We identified the need for a \textit{File Upload Exploitation Lab} (\FUEL) that unifies a diverse set of 15 UFU-specific scenarios into a single test framework, enabling a comprehensive and reproducible comparison of UFU scanners.

\subsection{Design}\label{sec:design}
A framework for the evaluation of UFU vulnerabilities and scanners should follow several design goals to make it adoptable in future work:
\begin{itemize}
    \item \textbf{Simplicity:} A simple setup is crucial to reduce friction for researchers to make it a go-to choice for their evaluation.
    \item \textbf{Extensiveness:} A diverse set of file upload vulnerability scenarios is necessary to evaluate the capabilities of vulnerability scanners. 
    \item \textbf{Extensibility:} A modular system that allows future work to extend the framework with new UFU scenarios.
\end{itemize}
\FUEL~achieves the first and third design goals with the use of containerization.
Docker \cite{docker} allows starting isolated containers agnostic to the operating system, thus facilitating the deployment and use of \FUEL.
Further, its \textit{compose} extension simplifies the orchestration of multiple containers with a single configuration file; therefore, deploying \FUEL~becomes a single command in its project directory: \texttt{docker compose up}

\begin{figure}[b]
    \centering
    \includegraphics[width=0.75\textwidth]{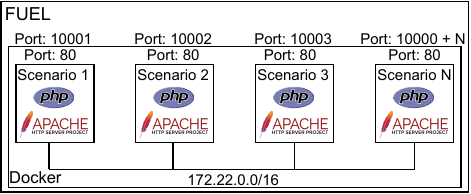}
    \caption{Depiction of \FUEL's design using multiple isolated containers.}
    \label{fig:fuel-design}
\end{figure}

Moreover, \FUEL~implements each UFU scenario in a separate container, each with its individual build instructions, as depicted in \cref{fig:fuel-design}.
Hence, each scenario can have an individual configuration of the web server and PHP environment and distinct source code for the vulnerable application.
All currently implemented scenarios in \FUEL~are based on PHP 8.3 and Apache httpd 2.4, with the web application being available from the host at port $10,000 + N$ or the internal network $172.22.0.N/16$.
Consequently, \FUEL~is trivial to extend with additional scenarios while offering file-system and process isolation of each web application to avoid interference, thus achieving the second goal.

\subsection{UFU Scenarios}
\FUEL~gives prominence to UFU vulnerabilities by implementing 15 distinct scenarios, which were distilled from related work, such as Pooj et al. \cite{pooj2016understanding}, OWASP \cite{owasp-ufu} or Portswigger \cite{portswigger-ufu}.
All scenarios have in common that files are uploaded to \texttt{/uploads/}, a directory inside the web server's document root, to make the files accessible.
While a brief description of each scenario follows, more detailed information is available in \FUEL's repository (cf. \cref{sec:open-science}).

\begin{enumerate}[label={S\arabic*}]
    \setlength\itemsep{0.47em}
    \item \textit{``Unrestricted''}:
This scenario poses as a baseline since no validation is performed on the uploaded file, allowing the upload of any file, including \texttt{.php} files.

    \item \textit{``Client-side''}:
This scenario only implements client-side validation.
By intercepting and manipulating the HTTP request, an attacker can circumvent the client-side validation to upload a PHP file.

    \item \textit{``MIME-type''}:
Only the uploaded file's MIME-type is validated server-side using the value provided in the \texttt{Content-Type} header.
The intended bypass is uploading a malicious file and changing the header to a permissible value, e.g., uploading \texttt{exploit.php} with type \texttt{image/png}.

    \item \textit{``Extension I''}:
Checking for known dangerous file extensions, such as \texttt{.php}, is modeled in this scenario.
Alternative extensions which are passed to the PHP interpreter, e.g., \texttt{.php5} or \texttt{.php8}, bypass this check.

    \item \textit{``Extension II''}:
Similar to the previous scenario, another bypass is changing the capitalization of individual file name's letters, e.g., \texttt{exploit.pHp}.

    \item \textit{``Extension III''}:
Misconfigured web servers might pass any file with \texttt{.php} in the file name, such as \texttt{exploit.php.png}, to the PHP interpreter, circumventing file extension checks.

    \item \textit{``Extension IV''}:
This scenario does not prevent the upload of configuration files, such as \texttt{.htaccess}, to change the web server's or PHP interpreter's behavior to execute malicious files.

    \item \textit{``Extension V''}:
Stripping prohibited extensions from file names is dangerous if not performed correctly, e.g., recursively.
The intended bypass for this scenario is nesting the file extension, such as \texttt{.p.phphp} becoming \texttt{.php}.

    \item \textit{``Magic number''}:
Determining the file type from the first or last bytes of the file content is insufficient, as the file's content may contain executable code.
For example, \texttt{GIF87a<?php phpinfo(); ?>} uploaded as \texttt{file.php} might be detected as a \texttt{image/gif} file due to the correct file header.

    \item \textit{``File content''}:
Similar to the previous scenario, valid image files may contain PHP code in their metadata.
Thus, checking the dimensions of an image file renamed to \texttt{file.php} succeeds, but so does the execution by the PHP interpreter when it encounters the PHP code.  

    \item \textit{``Directory configuration''}:
This scenario disables the PHP interpreter for the \texttt{uploads/} folder.
However, path traversal techniques may allow uploading in another location, such as \texttt{/}, where the PHP interpreter is not disabled.

    \item \textit{``Special characters''}:
Null bytes or other special characters may not be handled correctly by the application, allowing it to bypass validation checks.
The intended bypass is placing a Null byte in the file name, e.g., \texttt{file.php\%00.png}.

    \item \textit{``XSS''}:
If file names are displayed to the user without escaping special characters, they may be used for XSS attacks, although the application processes no file content.
This scenario models the upload of, e.g., \texttt{<img src=x onerror=alert(1)>.png}, which will trigger a JavaScript alert.

    \item \textit{``Race condition''}:
External programs, such as antivirus scanners, might delete or block access to uploaded files if found to be malicious.
Such an analysis usually incurs a delay, which might give attackers enough time to access and execute the file.
The intended exploit is uploading a file and accessing it before the web server returns the HTTP response after deleting it.

    \item \textit{``PUT method''}:
Web applications may use JavaScript or other technologies to dynamically perform HTTP requests to a backend server, including changes to the HTTP request method.
This scenario models a situation in which an HTTP PUT request is used to create and update files on the server.

\end{enumerate}

\section{Evaluation}\label{sec:evaluation}
While the related work presented in \cref{sec:related-work} showed that UFU vulnerabilities were extensively studied over the recent years, continuously improving detecting techniques and discovering previously unknown vulnerabilities, the following question motivated the upcoming evaluation:
\begin{itemize}
    \item RQ1: How capable are available UFU scanners against the file upload scenarios described in related work that we implemented in \FUEL?
\end{itemize}

\subsection{UFU Scanners}
For the evaluation, we review (non-)academic work for UFU scanners and identify several that do not require access to the web application's source code, e.g., do not perform static analysis.

To our surprise, \textit{FUSE} \cite{github-fuse,lee2020fuse} was the only publicly available academic, black-box UFU scanner.
For other publications implementing black-box vulnerability scanners, including \textit{UChecker} \cite{huang2019uchecker}, \textit{UFuzzer} \cite{huang2021ufuzzer}, \textit{FileUploadChecker} \cite{wichmann2022fileuploadchecker} or \textit{URadar} \cite{chenUradar2023}, we were unable to obtain source code.
However, at the time of writing, the latter already had a GitHub repository\footnote{\url{https://github.com/Cyc1e183/URadar}} without content.

In contrast, we discovered two open-source and one commercial scanner to include in our evaluation.
\textit{Fuxploider} \cite{fuxploider} is an open-source scanner specializing in flaws in file upload forms.
The second open-source scanner is \textit{ZAP} \cite{owasp-zap} in combination with the \textit{FileUpload} plugin \cite{owasp-zap-file-upload-addon}.
The commercial vulnerability scanner is \textit{BurpSuite Professional} \cite{portswigger-burp} with the \textit{UploadScanner} plugin \cite{upload-scanner}.

In general, all black-box UFU scanners follow a similar detection methodology.
First, a new variation of a potentially malicious file or HTTP request is created and sent to the server.
After the server responds, the scanner tries to access the uploaded file to identify successful XSS or CE exploits. 

\subsection{Methodology}

As all scenarios in \FUEL are isolated, we ran one UFU scanner against each scenario consecutively. 
Afterward, \FUEL~was reset to its initial state, and the next UFU scanner was tested.
A timeout of 300 seconds was set for each scenario, after which the scanning process was interrupted to continue with the next scenario.
Upon successful or forced termination, the scanner's output was analyzed for three aspects:
\begin{enumerate}
    \item Was the scanner able to use the intended file upload bypass (iFUB)? 
    \item Was the scanner able to gain and report code execution (CE)?
    \item Was the scanner able to gain and report cross-site scripting (XSS)?
\end{enumerate}

Each vulnerability scanner requires its own specific configuration. 
\textit{FUSE} and \textit{Fuxploider} are command-line scanners while \textit{ZAP} and \textit{BurpSuite} provide a graphical interface.
For example, the command-line scanners required a configuration file or a set of command-line arguments, e.g., specifying the upload endpoint and a detection criterion.
In contrast, the graphical scanners required a full HTTP request and a detection criterion.
We did two separate passes for the graphical scanners to initialize them with two different HTTP upload requests: A small image file (\texttt{fuel.png}) and a text file (\texttt{fuel.txt}) to avoid biases in the provided file format.
The command-line scanners did not require any file or full HTTP request to run.

In general, we aimed to keep default settings, configure as little as possible, and provide as little manual assistance as needed to keep the results unbiased and reproducible.
Naturally, we publish all details, such as configuration files, execution details, output logs, and software versions, to reproduce our results as part of our commitment to open science in our repository (cf. \cref{sec:open-science}).

The evaluation was conducted on a single system with 64 GB of RAM, an 8-core/16-thread CPU (AMD Ryzen 7 4800H), and a PCIe NVMe SSD drive. 

\subsection{Results}
\newcommand{\ok}{\textbf{\textcolor{blue}{\checkmark}}}
\newcommand{\gok}{\textbf{\textcolor{gray}{\checkmark$^1$}}}
\newcommand{\nok}{\textcolor{red}{$\times$}}
\newcommand{\gnok}{\textcolor{gray}{$\times$}}
\newcommand{\timeout}{\textcolor{gray}{T}}
All scanners required less than 300 seconds to finish their analysis, except in scenario 14.
\Cref{tab:allres} shows the results of each UFU vulnerability scanner executed against \FUEL's scenarios.
Scenario 13 did not feature a way to achieve code execution, as indicated by \texttt{-}.
Similarly, \textit{Fuxploider} does not implement XSS detection techniques, leaving the column unmarked.

\subsubsection{Cross-Site Scripting}
The data shows that each scanner discovered more XSS than CE vulnerabilities.
\FUEL's current scenarios emphasize UFU vulnerabilities with a high impact, such as code execution, therefore not actively preventing the upload of files containing XSS payloads.  Nonetheless, we include XSS for completeness in \cref{tab:allres}.

\subsubsection{(Un)exploited Scenarios}
Notably, all scanners successfully discovered the vulnerabilities in scenarios 1 - 4 if initialized with an image file.
Otherwise, ZAP's FileUpload plugin failed to identify any vulnerability in scenario 3, and BurpSuite's UploadScanner did not recognize the XSS.
Nonetheless, from scenario 5 on, the detection capabilities began to differ.
No scanner achieved a successful exploit against the file upload mechanism to gain code execution in scenarios 8, 11, 14, and 15, which require file extension manipulation, path traversal, race condition, and request manipulation capabilities.

\subsubsection{Scanner Capabilities}
\textit{FUSE} implements UFU exploitation techniques for 8 out of 15 scenarios, detecting a total of 7 CE and 11 XSS vulnerabilities.
It is the only scanner not to exploit CE in scenario 6, while struggling with scenarios 8 and 10 through 15.
Surprisingly, it detects the iFUB in scenario 7 but does not exploit the resulting CE.

\begin{table}[t]
\setlength{\tabcolsep}{3pt}
\caption{Results from running each UFU scanner against \texttt{FUEL}'s scenarios.}
\label{tab:allres}
\centering
\begin{tabular}{c*{12}{c}} 
\toprule
Scanner & \multicolumn{3}{c}{\textbf{FUSE}} & \multicolumn{3}{c}{\textbf{Fuxploider}} & \multicolumn{3}{c}{\textbf{ZAP}} & \multicolumn{3}{c}{\textbf{BurpSuite}}\\
\cmidrule(r){2-4} \cmidrule(lr){5-7} \cmidrule(lr){8-10} \cmidrule(l){11-13}
Type & iFUB & CE & XSS & iFUB & CE & XSS & iFUB & CE & XSS & iFUB & CE & XSS\\
\midrule
Total & 8 & 7 & 11 & 8 & 8 & - & 9 & 8 & 11 & 9 & 8 & 12\\
\midrule
S1   & \ok & \ok & \ok & \ok & \ok & - & \ok & \ok & \ok & \ok & \ok & \ok\\
S2   & \ok & \ok & \ok & \ok & \ok & - & \ok & \ok & \ok & \ok & \ok & \ok\\
S3   & \ok & \ok & \ok & \ok & \ok & - & \gok & \gok & \gok & \ok & \ok & \gok\\
S4   & \ok & \ok & \ok & \ok & \ok & - & \ok & \ok & \ok & \ok & \ok & \ok\\
S5   & \ok & \ok & \ok & \ok & \ok & - & \ok & \ok & \ok & \nok & \nok & \ok\\
S6   & \nok & \nok & \ok & \ok & \ok & - & \ok & \ok & \ok & \ok & \ok & \ok\\
S7   & \ok & \nok & \ok & \ok & \ok & - & \nok & \nok & \ok & \nok & \nok & \ok \\
S8   & \nok & \nok & \ok & \nok & \nok & - & \nok & \nok & \ok & \nok & \nok & \ok\\
S9   & \ok & \ok & \ok & \nok & \nok & - & \ok & \ok & \nok & \ok & \ok & \ok\\
S10  & \ok & \ok & \ok & \nok & \nok & - & \nok & \nok & \nok & \ok & \ok & \ok\\
S11  & \nok & \nok & \ok & \nok & \nok & - & \nok & \nok & \ok & \nok & \nok & \ok\\
S12  & \nok & \nok & \nok & \ok & \ok & - & \gok & \gok & \gok & \ok & \ok & \nok\\
S13  & \nok & - & \nok & - & - & - & \ok & - & \ok & \ok & - & \ok\\
S14  & \timeout & \timeout & \timeout  & \nok & \nok & - & \timeout & \timeout & \timeout & \timeout & \timeout & \timeout \\
S15  & \nok & \nok & \nok & \nok & \nok & - & \nok & \nok & \nok & \nok & \nok & \nok\\ 
\bottomrule
\end{tabular}
\\ \ok: Found, \gok: Only for \texttt{fuel.png}, \nok: Not found, \timeout: Timeout after 300s 
\end{table}

Although \textit{Fuxploider} lacks XSS capabilities, it always gains CE after identifying a UFU, which it achieves in 8 different scenarios.
It did not identify the iFUB that requires file header or content manipulation, race condition, or path traversal capabilities.
However, it is the only scanner to successfully exploit the CE in scenario 7.

The UFU vulnerability scanning plugin \textit{FileUpload} for ZAP detects 9 of 15 iFUBs and exploits 8 out of 14 possible CE vulnerabilities but fails to identify the vulnerabilities in scenarios 7, 8, 10, and 11. 
However, it only achieves these comparatively good results when initialized with an image file.
If it is initialized with a text file, it fails to find the vulnerabilities in scenarios 3 and 12.

While BurpSuite Professional's \textit{UploadScanner} plugin discovers the most XSS vulnerabilities and 9 out of 15 iFUBs, it is the only scanner not to change the capitalization of a file's extension (scenario 5).
Moreover, it fails to bypass the checks in scenarios 7, 8, and 11, which require uploading configuration files, nesting file extensions, or performing path traversal.
Similar to the FileUpload plugin, the UploadScanner's results change if initialized with a non-image file, although it only affects the XSS detection in scenario 3.

\subsection{Evaluating our Fuxploider-NG}
With the help of \FUEL, we have identified several shortcomings in existing UFU scanners, as no scanner has the capabilities necessary to exploit all 15 scenarios.
To demonstrate \FUEL's~effectiveness in improving existing work, we develop \textit{Fuxploider-NG} by extending Fuxploider with additional UFU exploitation and XSS detection capabilities.
In particular, we implemented placing PHP code after magic bytes or inside an image's metadata and using nested file extensions, path traversal, or race condition techniques.

\Cref{tab:fuxploider-ng} shows the results from testing \textit{Fuxploider-NG} with its improvements and updated configuration against \FUEL. 
The results indicate that it successfully discovers all iFUBs, CE, and XSS vulnerabilities in scenarios 1 through 14.
Only the last scenario could not be improved as it would require major changes to \textit{Fuxploider}'s code base and is therefore considered future work.

\begin{table}[b]
\setlength{\tabcolsep}{4pt}
\centering
\caption{Our improved \textit{Fuxploider-NG} scanner successfully exploits scenarios 1 - 14.}
\label{tab:fuxploider-ng}
\begin{tabular}{@{} c *{15}{c} @{}}
\toprule
& S1 & S2 & S3 & S4 & S5 & S6 & S7 & S8 & S9 & S10 & S11 & S12 & S13 & S14 & S15 \\
\midrule
iFUB & \ok & \ok & \ok & \ok & \ok & \ok & \ok & \ok & \ok & \ok & \ok & \ok & \ok & \ok & \nok \\
CE   & \ok & \ok & \ok & \ok & \ok & \ok & \ok & \ok & \ok & \ok & \ok & \ok & -   & \ok & \nok \\
XSS  & \ok & \ok & \ok & \ok & \ok & \ok & \ok & \ok & \ok & \ok & \ok & \ok & \ok & \ok & \nok \\
\bottomrule
\end{tabular}
\end{table}

\section{Discussion}
This section answers the research question RQ1 by interpreting the data from \cref{tab:allres} and discussing the evaluated scanners' shortcomings.
Then, it discusses the results from Fuxploider-NG and the importance of our \FUEL~framework.
Finally, the section closes with limitations of \FUEL~and possible future work in the UFU domain.

\subsection{Shortcomings of UFU Vulnerability Scanners}
Albeit all scanners exploited the first two UFU scenarios, the results show that the UFU detection capabilities are inhomogeneous, as no scanner managed to exploit all of \FUEL's scenarios.
Exploitation of the first two scenarios by all scanners is unsurprising since no server-side validation is performed, and any malicious file can be uploaded.

To our surprise, ZAP's and BurpSuite's plugins have difficulties finding the vulnerabilities in scenario 3 if initialized with a text file but find them when initialized with an image file.
One reason for ZAP's difference is assumed to be the missing manipulation of the MIME-type in the HTTP request.

Nonetheless, all scanners know to check for alternative, executable PHP extensions (scenario 4) and, except for Burp's plugin, change the extension's capitalization (scenario 5).
Reuse of scenario 4's exploit in scenario 5 is apparent but impossible due to scenario 5's web server configuration.
Although file extension bypasses appear well studied and implemented, FUSE is the only scanner not to exploit scenario 6, which requires an infix \texttt{.php} extension.

On the contrary, Fuxploider is the only scanner to exploit scenario 7 successfully by uploading an \texttt{.htaccess} configuration file to gain code execution.
Although FUSE detects the possibility of uploading such files, it does not actively exploit it.
Scenario 8 removes the string \texttt{.php} from the file's name to model an insecure attempt to mitigate dangerous extensions.
According to the results, either no scanner has implemented such file name variations, or the exploit validation fails by trying to access the original file name but not the modified one. 

Scenario 9 required a file with a valid image file header (e.g., \texttt{GIF87a} followed by PHP code) to bypass the upload filter and gain CE.
Except for Fuxploider, all scanners successfully performed this iFUB. 
A more advanced version of this exploit is required for scenario 10, as the file's content must be valid image data, and the malicious PHP code must be placed in the image's metadata.
This check tripped ZAP's plugin, so only FUSE and Burp's UploadScanner successfully gained CE.

No scanner successfully uses exploitation methods that rely on path traversal by manipulating the file name, as required for scenario 11.
During this evaluation, all scanners knew that uploaded files would be stored in \texttt{/uploads/}, but this might not be the case for other web applications; thus, rediscovering the uploaded file becomes a difficult challenge. 
Even more so if the new file path leaves the document root or the accessible directory.
Thus, such an exploitation method might have little chance of success but should be implemented anyhow, as demonstrated in Fuxploider-NG. 

Except for FUSE, all scanners use special characters to gain code execution in scenario 12. 
Interestingly, ZAP's plugin only successfully exploits this scenario if initialized with an image file.
We assume that while the plugin inserts special characters into the file name, it only manipulates the provided file name from the initialized HTTP request and thus does not change \texttt{fuel.txt} to \texttt{fuel.txt\%00.png}, which would exploit this scenario.
In FUSE's case, its authors might have assumed this bypass to be obsolete, as FUSE was published in 2020 and PHP handles Null bytes correctly since version 5.3 (released in 2011 \cite{php-nullbytes}).
Nonetheless, we included this scenario in \FUEL~and Fuxploider-NG for completeness and to match the capabilities of BurpSuite's plugin.

Scenario 13 does not process or store the file, except for displaying its name without escaping, making it susceptible to XSS.
While this is not directly a UFU, it is related and may become a severe security issue, even if all other file properties are correctly validated.
Still, only 50\% of the scanners are capable of detecting this issue, while FUSE and Fuxploider lack the capabilities.

Scenario 14's validation implements a processing delay of 0.5 seconds and deletes the file before returning the HTTP response, which confuses all scanners.
One possible explanation is strictly following the pattern of first uploading, then verifying, and waiting for a successful HTTP response to the upload request before attempting to access the file.
In Fuxploider-NG, we implemented additional access attempts right after the upload without waiting for the HTTP response and thus successfully exploited this scenario.

Although we expected the scanners not to discover the iFUB in scenario 15, we still decided to include it for completeness since it was mentioned in related work.
An explanation for the results is that the PUT upload requests are sent using JavaScript, and no scanner features functionality to successfully extract the request parameters or does not implement mutations of the HTTP method and request URL.
Therefore, a longer timeout would not have changed the results.

\subsection{Comparison to Fuxploider-NG}

Due to the various shortcomings, it is not possible to determine the best-performing UFU scanner, as the capabilities differ greatly.
Even if the unexploited scenarios 8,11,14, and 15 are not considered, each scanner fails at exploiting at least two scenarios.
Consequently, one would need to use at least two scanners to evaluate a web application for a set cover of the exploited UFU scenarios. 
The three possible minimal combinations to cover scenarios 1-7,9-10, and 12-13 are FUSE and ZAP's plugin initialized with an image, FUSE and BurpSuite's plugin, Fuxploider and BurpSuite's plugin.

As BurpSuite is a commercial tool with a subscription model costing several hundred Euros per year, FUSE and ZAP are the only free-to-use combination. 
However, in that case, the user must be aware that the ZAP plugin's accuracy depends on the initialization of the HTTP request, which is not documented in its repository.
Furthermore, the need to use at least two scanners increases the friction due to the configurations' complexity and multiplies the testing duration, as most tests will be performed twice.

With Fuxploider-NG we contribute a single scanner that covers 14 out of 15 scenarios currently present in \FUEL~and successfully exploits them, thus reducing the complexity of testing a web application for UFU vulnerabilities.

\subsection{Importance of \FUEL}
Using \FUEL, we have shown that the capabilities of academic, open-source, and commercial vulnerability scanners are limited.
Each vulnerability scanner has a different set of UFU exploitation techniques, thus leaving room for improvement to detect the 15 types of UFU vulnerabilities.
This result is quite surprising, as related work has discussed these vulnerability scenarios since 2016, reaffirming the need for a standardized evaluation framework such as \FUEL.

Consequently, using any of these scanners against real-world applications to assess their security entails the risk of false negatives, thus leaving its users with a false sense of security and web applications with a potential, undiscovered entry point for malicious actors.

A collection of tests to ensure new code does not break existing functionality is a core concept of test-driven development (TDD) \cite{tdd}.
Parallels can be drawn between a test suite of TDD and the set of UFU scenarios in \FUEL, thus making \FUEL~an important framework in ensuring future work in the UFU domain covers all previously documented vulnerable scenarios in its goal to discover novel techniques.

Furthermore, \FUEL's modularity embraces adding new scenarios, allowing the modeling of other UFU vulnerabilities.
Finally, \FUEL~facilitates a fair, reliable, and reproducible way to evaluate improvements against related work.

\subsection{Limitations \& Future work}

For now, \FUEL's scenarios solely focus on UFU vulnerabilities in PHP. 
While many websites use PHP, other languages are suitable for web development and potentially vulnerable to UFU, too, although the exploitation methods might differ. 
For example, not invoking an interpreter for each request might not directly execute malicious code inside a file.
Since related work and the examined UFU scanners primarily focus on PHP, it naturally followed to implement \FUEL's scenarios in PHP.
However, future work can implement scenarios in other languages, as \FUEL~is language agnostic.

Each of the 15 scenarios was deliberately kept simple by modeling exactly one validation check.
While this approach allows to efficiently evaluate UFU scanners, real-world applications are likely to perform multiple checks.
However, \FUEL~is not limited to simple scenarios, as future work could implement scenarios based on real-world applications or combinations of existing scenarios.

Although we noticed variations in the scanners' performance, e.g., the number of requests or execution time, an analysis was considered future work for the following reasons.
First, some scanners did not report these metrics or showed inconsistent runtime behavior.
Second, it raises new questions, such as the trade-off between a faster scanner with less accuracy and a slower but more thorough and accurate scanning technique.
Furthermore, some scanners terminated after the first finding, while others continued to examine the application.

\subsection{Open Science}\label{sec:open-science}
It is unfortunate that several UFU scanners from related work were unavailable for this evaluation, thus limiting it to only four scanners.
As the authors share strong beliefs in open and reproducible science, we will open-source \FUEL\footnote{\url{https://github.com/FUEL-Project/FUEL-FileUploadExploitationLab}}, Fuxploider-NG\footnote{\url{https://github.com/FUEL-Project/fuxploider-NG}}, and all results related to our evaluation\footnote{\url{https://github.com/FUEL-Project/FUEL-Evaluation}} on GitHub. 
Notably, we encourage and welcome contributions to \FUEL, especially new scenarios, to make it a comprehensive evaluation suite for UFUs.

\section{Conclusion}
Our review of related work revealed a lack of a standardized evaluation methodology for UFU vulnerability scanners, which could be one possible reason for the divergence in detection capabilities of the evaluated academic, open source, and commercial vulnerability scanners. 
Thus, our first contribution is a novel framework, which we call \FUEL, consisting of 15 distinct file upload scenarios distilled from related work. 

The framework offers a standardized environment to facilitate a comprehensive and reproducible evaluation of existing vulnerability scanners, new UFU detection methods, or new vulnerable file-handling scenarios.
Furthermore, \FUEL's design goals were simplicity, extensibility, and reproducibility to make this framework suitable for future work in the unrestricted file upload vulnerability domain. 

With the help of \FUEL, our results show that none of the evaluated black-box UFU scanners is capable of detecting all previously documented UFU scenarios. 
Each scanner has individual shortcomings, requiring users to test a web application with at least two scanners for a comprehensive analysis or leaving them with a false sense of security.

We propose \textit{Fuxploider-NG}, which consolidates all UFU vulnerability detection methods into a single program, thus closing the capability gap and reducing the risk of false negatives when testing real-world applications for UFU vulnerabilities.

Finally, by open-sourcing \FUEL~and Fuxploider-NG, we aim to improve existing vulnerability scanners and lay the groundwork for more research to increase the number of detected UFU vulnerabilities in web applications.

\vfill
\bibliographystyle{splncs04}
\bibliography{bibliography}

\end{document}